\documentstyle[floats,prd,aps,epsfig,eqsecnum,12pt]{revtex}
\makeatletter
\newbox\tempboxa
\newdimen\captionboxsubcount
\def\capsize#1{\captionboxsubcount=#1pt}
\newdimen\captionboxsub
\captionboxsub=\hsize \advance\captionboxsub by -\captionboxsubcount
\advance\captionboxsub by -\captionboxsubcount
\long\def\@makecaption#1#2{
 \setbox\@tempboxa\hbox{#1 #2}
 \ifdim \wd\@tempboxa >\captionboxsub
\rightskip=\captionboxsubcount \leftskip=\captionboxsubcount #1 #2
\else \hbox to\hsize{\hfil\box\@tempboxa\hfil}
 \fi}
\makeatother
\capsize{30}

\begin{document}
\begin{titlepage}
\begin{flushright}
\begin{minipage}{5cm}
\begin{flushleft}
\small
\baselineskip = 10pt
YCTP-P-29-99
\end{flushleft}
\end{minipage}
\end{flushright}
\begin{center}
\Large\bf
 Phases of Chiral Gauge Theories
\end{center}
\footnotesep = 12pt
\vfill
\begin{center}
\large
\centerline{Thomas {\sc Appelquist}\footnote{Electronic address: {\tt
thomas.appelquist@yale.edu}}  \quad Zhiyong {\sc
Duan}\footnote{Electronic address: {\tt zhiyong.duan@yale.edu}}
\quad Francesco {\sc Sannino} \footnote{Electronic address : {\tt
francesco.sannino@yale.edu}}}
\vskip 1cm
 {\it Department of Physics, Yale University, New
Haven,\\~CT~06520-8120,~USA.}
\end{center}
\vfill
\begin{center}
\bf Abstract
\end{center}
\begin{abstract}
\baselineskip = 17pt
We discuss the behavior of two non-supersymmetric chiral $SU(N)$
gauge theories, involving fermions in the symmetric and
antisymmetric two-index tensor representations respectively. In
addition to global anomaly matching, we employ a recently proposed
inequality constraint on the number of effective low energy
(massless) degrees of freedom of a theory, based on the
thermodynamic free energy. Several possible zero temperature phases
are consistent with the constraints. A simple picture for the phase
structure emerges if these theories choose the phase, consistent
with global anomaly matching, that minimizes the massless degree of
freedom count defined through the free energy. This idea suggests
that confinement with the preservation of the global symmetries
through the formation of massless composite fermions is in general
not preferred. While our discussion is restricted mainly to
bilinear condensate formation, higher dimensional condensates are
considered for one case. We conclude by commenting briefly on two
related supersymmetric chiral theories.

\end{abstract}
\begin{flushleft}
\end{flushleft}
\vfill
\end{titlepage}

\section{Introduction}
\label{uno}
Chiral gauge theories, in which at least part of the matter field
content is in complex representations of the gauge group, play an
important role in efforts to extend the standard model. These
include grand unified theories, dynamical breaking of symmetries,
and theories of quark and lepton substructure. An important
distinction from vector-like theories such as QCD is that since at
least some of the chiral symmetries are gauged, mass terms that
would explicitly break these chiral symmetries are forbidden in the
Lagrangian. Another key feature is that the fermion content is
subject to a constraint not present in vectorial gauge theories,
the cancellation of gauge and gravitational anomalies.

Chiral theories received much attention in the 1980's~\cite{ball},
focusing on their strong coupling behavior in the infrared. One
possibility is confinement with the gauge symmetry as well as
global symmetries unbroken, realized by the formation of gauge
singlet, massless composite fermions. Another is confinement with
intact gauge symmetry but with some of the global symmetries broken
spontaneously, leading to the formation of gauge-singlet Goldstone
bosons. It is also possible for these theories to exist in the
Higgs phase, dynamically breaking their own gauge
symmetries~\cite{rds}. Depending on particle content, they might
even remain weakly coupled. This will happen if the theory has an
interacting but weak infrared fixed point. The symmetries will then
remain unbroken, and the infrared and underlying degrees of freedom
will be the same.

Supersymmetric (SUSY) chiral theories have also received
considerable attention over the years, since most of the known
examples of dynamical supersymmetry breaking involve these kinds of
theories \cite{pt}.

 Studies of chiral gauge theories have typically made use of
the 't Hooft global anomaly matching conditions~\cite{thooft} along
with $1/N$ expansion, and not-so-reliable most attractive channel
(MAC) \cite{peskin} analysis and instanton computations. Direct
approaches using strong coupling lattice methods \cite{92conf} are
still difficult. Another indirect approach developed recently
\cite{acs} takes the form of an inequality limiting the number of
massless degrees of freedom in the infrared description of a field
theory to be no larger than the number of ultraviolet degrees of
freedom. It was conjectured to apply to all asymptotically free
theories whose infrared behavior is also governed by a fixed point,
not necessarily free.

The inequality is formulated using finite temperature as a device to probe
all energy scales, with the degree-of-freedom count defined using the free
energy of the field theory. The zero-temperature theory of interest is
characterized using the quantity $f_{IR}$, related to the free energy by
\begin{equation}  \label{eq:firdef}
f_{IR} \equiv - \lim_{T\to 0} \frac{{\cal F}(T)}{T^4}
\frac{90}{\pi^2} \ ,
\end{equation}
where $T$ is the temperature and ${\cal F}$ is the conventionally
defined free energy per unit volume. The limit is well defined if
the theory has an infrared fixed point. For the special case of an
infrared-free theory, $f_{IR}$ is simply the number of massless
bosons plus $7/4$ times the number of 2-component massless Weyl
fermions. The corresponding expression in the large $T$ limit is
\begin{equation}  \label{eq:fuvdef}
f_{UV} \equiv - \lim_{T\to \infty} \frac{{\cal F}(T)}{T^4}
\frac{90}{\pi^2}\ .
\end{equation}
This limit will be well defined if the theory has an ultraviolet
fixed point. For an asymptotically free theory $f_{UV}$ counts the
underlying, ultraviolet degrees of freedom in a similar way.

In terms of these quantities, the conjectured inequality for any
asymptotically-free theory is
\begin{equation}  \label{eq:ineq}
f_{IR} \le f_{UV}\ .
\end{equation}
This inequality has not been proven, but in Ref.~\cite{acs} it was
shown to agree with all known results and then used to derive new
results for several strongly coupled, vector-like gauge theories.
It was applied to chiral theories in Ref.~\cite{acss}. The
principal focus there was on the possibility of preserving the
global symmetries through the formation of massless composite
fermions.

In this paper, we examine further the two non-supersymmetric chiral
theories of Ref.~\cite{acss}, both rich in possible phase
structure. One is the Bars-Yankielowicz (BY) \cite{by} model
involving fermions in the two-index symmetric tensor
representation, and the other is a generalized Georgi-Glashow (GGG)
model involving fermions in the two-index antisymmetric tensor
representation. In each case, in addition to fermions in complex
representations, a set of $p$ anti fundamental-fundamental pairs
are included and the allowed phases are considered as a function of
$p$. Several possible phases emerge, consistent with global anomaly
matching and the above inequality.

Graphs for the various $f_{IR}$'s vs $p$ are plotted for each
model. They lead us to the suggestion that each of these theories
will choose from among the allowed phases the one that minimizes
$f_{IR}$ at each value of $p$. This idea is useful only to
distinguish among phases for which the $f_{IR}$'s are computable,
those that are infrared free or governed by a weak infrared fixed
point. It may be rephrased in terms of the entropy per unit volume
$S(T)$ of the system (the derivative of the pressure per unit
volume $P(T)
= -{\cal F}(T)$). Provided only that the limit in
Eq.~(\ref{eq:firdef}) exists, $S(T)$ near freeze out will be given
by $S(T)
= (2\pi^{2}/45) T^{3} f_{IR}$ plus higher order terms in the low
temperature expansion. Thus the minimum number of degrees of
freedom corresponds to the minimum entropy at approach to freeze
out, consistent with global anomaly matching. Now of course at any
finite T, the preferred phase is chosen from among all the states
entering the partition function by minimizing the free energy
density, which becomes the energy density at $T=0$. Comparing these
quantities for different states when the theory is strongly
interacting is, however, generally a strong coupling problem. The
conjecture for these theories is the following. Assuming that the
candidate phases revealed by anomaly matching {\it are} states
entering the partition function, the state with the relatively
lower energy density is approached at a lower rate proportional to
the lower $f_{IR}$. The latter quantity is computable in terms of
only the effective low energy theory which is infrared free for the
phases being considered here.

We focus almost completely on symmetry breaking patterns
corresponding to the formation of bilinear condensates. We suggest
that in general, the phase corresponding to confinement with all
symmetries unbroken, where all the global anomalies are matched by
massless composite fermions, is not preferred. Instead, the global
symmetries associated with fermions in real representations break
spontaneously via bilinear condensate formation as in QCD. With
respect to the fermions in complex representations, however, the
formation of bilinear condensates is suggested to be disfavored
relative to confinement and the preservation of the global
symmetries via massless composite fermion formation.

It is interesting to note that in the real world case of two-flavor
QCD (a vector-like theory with {\it all} fermions in a real
representation) nature prefers to minimize $f_{IR}$. Neglecting the
small bare quark masses, global anomaly matching admits two
possible low energy phases, broken chiral symmetry through the
formation of the bilinear $<{\bar\psi}\psi
>$ condensate, or unbroken chiral symmetry through the formation of confined
massless baryons. Both effective low energy theories are infrared
free. The three Goldstone bosons of the former (chosen by nature)
lead to $ f_{IR}= 3$, and the two massless composite Dirac fermions
of the latter lead to $f_{IR}= 7$.

Bilinear condensate formation is of course not the only possibility
in a strongly coupled gauge field theory. We have so far extended
our discussion to include general condensate formation for one
simple example, the $SU(5)$ Georgi-Glashow model, which has
fermions in only complex representations and has only a $U(1)$
global symmetry. This symmetry can be broken via only higher
dimensional condensates. Interestingly, this breaking pattern, with
confinement and unbroken gauge symmetry, leads to the minimum value
of $f_{IR}$. This highlights the important general question of the
pattern of symmetry breaking in chiral theories when arbitrary
condensate formation is considered. Higher dimensional condensates
might play an important role, for example, in the dynamical
breaking of symmetries in extensions of the standard model
\cite{HDC}.

The Bars-Yankielowicz model is discussed in Section \ref{due} and
the generalized Georgi-Glashow model is discussed in Section
\ref{tre}. In Section \ref{quattro}, we briefly describe two
supersymmetric chiral theories: the supersymmetric version of the
$SU(5)$ Georgi-Glashow model and the closely related $3-2$ model.
In Section \ref{cinque} we summarize and conclude.

\section{ The Bars Yankielowicz (BY) Model}
\label{due}

This model is based on the single gauge group $SU(N\geq 3) $ and
includes fermions transforming as a symmetric tensor
representation, $S=\psi
_{L}^{\{ab\}}$, $a,b=1,\cdots ,N$; $\ N+4+p$ conjugate fundamental
representations: $\bar{F}_{a,i}=\psi _{a,iL}^{c}$, where $i=1,\cdots ,N+4+p$%
; and $p$ fundamental representations, $F^{a,i}=\psi _{L}^{a,i},\ i=1,\cdots
,p$. The $p=0$ theory is the basic chiral theory, free of gauge
anomalies by virtue of cancellation between the antisymmetric
tensor and the $N+4$ conjugate fundamentals. The additional $p$
pairs of fundamentals and conjugate fundamentals, in a real
representation of the gauge group, lead to no gauge anomalies.

The global symmetry group is
\begin{equation}
G_{f}=SU(N+4+p)\times SU(p)\times U_{1}(1)\times U_{2}(1)\ .
\label{gglobal3}
\end{equation}
Two $U(1)$'s are the linear combination of the original $U(1)$'s generated
by $S\rightarrow e^{i\theta _{S}}S$ , $\bar{F}\rightarrow e^{i\theta _{\bar{F%
}}}\bar{F}$ and $F\rightarrow e^{i\theta _{F}}F$ that are left invariant by
instantons, namely that for which $\sum_{j}N_{R_{j}}T(R_{j})Q_{R_{j}}=0$,
where $Q_{R_{j}}$ is the $U(1)$ charge of $R_{j}$ and $N_{R_{j}}$ denotes
the number of copies of $R_{j}$.

Thus the fermionic content of the theory is
\begin{equation}
\begin{tabular}{cccccc}
& $[SU(N)]$ & $SU(N+4+p)$ & $SU(p)$ & $U_{1}(1)$ & $U_{2}(1)$ \\
&  &  &  &  &  \\
$S$ &
\begin{tabular}{|c|c|}
\hline
$\ \ $&$\ \ $  \\ \hline
\end{tabular}
& $1$ & $1$ & $N+4$ & $2p$ \\ &  &  &  &  &  \\ $\bar{F}$ &
$\overline{
\begin{tabular}{|c|}
\hline
$\ \ $\\ \hline
\end{tabular}
}$ & $\overline{
\begin{tabular}{|c|}
\hline
$\ \ $\\ \hline
\end{tabular}
}$ & $1$ & $-(N+2)$ & $-p$ \\
&  &  &  &  &  \\
$F$ &
\begin{tabular}{|c|}
\hline
$\ \ $\\ \hline
\end{tabular}
& $1$ &
\begin{tabular}{|c|}
\hline
$\ \ $\\ \hline
\end{tabular}
& $N+2$ & $-(N-p)$
\end{tabular}
\end{equation}
where the first $SU(N)$ is the gauge group, indicated by the square brackets.

For all the models considered in this paper, the beta function is
generically written as
\begin{equation}
\beta =\mu \frac{d\alpha }{d\mu }=-\beta _{1}{\LARGE (}\frac{\alpha ^{2}}{
2\pi }{\Large )}-\beta _{2}{\large (}\frac{\alpha ^{3}}{4\pi
^{2}}{\large )} +O(\alpha ^{4}) \ , \label{betafun}
\end{equation}
where the terms of order $\alpha ^{4}$ and higher are scheme-dependent. For
the present model, we have $\beta _{1}=3N-2-(2/3)p$ and $\beta
_{2}=(1/4)\{13N^{2}-30N+1+12/N-2p((13/3)N-1/N)\}$. Thus the theory is
asymptotically free for
\begin{equation}
p<(9/2)N-3 \ . \label{afby}
\end{equation}
We shall restrict $p$ so that this condition is satisfied.

Because of asymptotic freedom, the thermodynamic free-energy may be computed
in the $T\rightarrow \infty $ limit. An enumeration of the degrees of
freedom leads to
\begin{equation}
f_{UV}=2(N^{2}-1)+\frac{7}{4}{\large
[}\frac{N(N+1)}{2}+(N+4)N+2pN{\Large ]} \ .
\label{finfby}
\end{equation}

The infrared realization of this theory will vary depending on the number $p$
of conjugate fundamental-fundamental pairs. We begin by discussing the $p=0$
theory and then map out the phase structure as function of $p$.

\subsection{The $p=0$ Case}

For $p=0$, the fermions are in complex representations of the
$SU(N)$ gauge group and the global symmetry group is
$G_{f}=SU(N+4)\times U_{1}(1)$. The theory is strongly coupled at
low energies, so it is expected either to confine or to break some
of the symmetries, consistent with global anomaly matching.

The possibility that the $p=0$ theory confines with the full global
symmetry group $G_{f}$ unbroken has been considered previously in
the literature~\cite{acss,eppz}. All the global anomalies of the
underlying theory may be matched at low energies providing that the
massless spectrum is composed of gauge singlet composite fermions
transforming according to the antisymmetric second-rank tensor
representation of $SU(N+4)$. They are described by the composite
operators $\bar{F}_{[i}S\bar{F}_{j]}$ and have charge $-N$ under
the $U_{1}(1)$ global symmetry.

With only these massless composites in the low energy spectrum, there are no
dimension-four interactions, so the composites are noninteracting in the
infrared. Therefore the thermodynamic free energy may be computed in the
limit $T\rightarrow 0$. Enumerating the degrees of freedom gives
\begin{equation}
f_{IR}^{sym}(p=0) = \frac{7}{4}\frac{(N+4)(N+3)}{2}\ ,
\label{fzerosym}
\end{equation}
where the superscript indicates that the full global symmetry is
intact. Clearly $f_{IR}^{sym}(p=0)< f_{UV}(p=0)$, satisfying the
inequality of Eq.~(\ref{eq:ineq})~\cite{acss}.

While the formation of confined massless composite fermions and the
preservation of $G_{f}$ is consistent with anomaly matching and the
thermal inequality, the same can be seen to be true of broken
symmetry channels. We consider first the Higgs phase corresponding
to the maximally attractive channel. It is~\cite{peskin}
\[
\begin{tabular}{|c|c|}
\hline
$\ \ $&$\ \ $  \\ \hline
\end{tabular}
\times \overline{
\begin{tabular}{|c|}
\hline
$\ \ $ \\ \hline
\end{tabular}
}\rightarrow
\begin{tabular}{|c|}
\hline
$\ \ $\\ \hline
\end{tabular}
~,
\]
leading to the formation of the $S\bar{F}$ condensate
\begin{equation}
\varepsilon ^{\gamma \delta }S_{\gamma }^{ai}\bar{F}_{\{a,i\},\delta }\ ,
\label{gauge breaking}
\end{equation}
where $\gamma ,\delta =1,2$ are spin indices and $a,i=1,\cdots ,N$
are gauge and flavor indices. This condensate breaks $U_{1}(1)$ and
all the gauge symmetries, and it breaks $SU(N+4)$ to $SU(4)$. But
the $SU(N)$ subgroup of $SU(N+4)$ combines with the gauge group,
leading to a new global symmetry $SU^{\prime }(N)$. For this group,
$\bar{F}_{a,i\leq N}$ is reducible, to the symmetric
$\bar{F}^{S}=\bar{F}_{\{a,i\}}$ and the anti-symmetric
$\bar{F}^{A}=\bar{F}_{[a,i]}$ representations.

The broken $SU(N+4)$ generator
\[
Q_{(N+4)}=\left(
\begin{tabular}{ccc|ccc}
$4$ &  &  &  &  &  \\ & $\ddots $ &  &  &  &  \\ &  & $4$ &  &  &
\\ \hline &  &  & $-N$ &  &  \\ &  &  &  & $\ddots $ &  \\ &  &  &
&  & $-N$
\end{tabular}
\right) \ ,
\]
combines with $Q_{1}$ giving a residual global symmetry $U_{1}^{\prime }(1)=$
$\frac{1}{N+4}(2Q_{1}-Q_{(N+4)})$. The breakdown pattern thus is
\begin{equation}
\left[ SU(N)\right] \times SU(N+4)\times U_{1}(1)\rightarrow SU^{\prime
}(N)\times SU(4)\times U_{1}^{\prime }(1) \ .
\label{rim}
\end{equation}

The gauge bosons have become massive as have some fermions. The fermionic
spectrum, with respect to the residual global symmetry is
\begin{equation}
\begin{tabular}{c|cccc}
&  & $SU^{\prime }(N)$ & $SU(4)$ & $U_{1}^{\prime }(1)$ \\
&  &  &  &  \\
& $S$ &
\begin{tabular}{|c|c|}
\hline
$\ \ $&$\ \ $  \\ \hline
\end{tabular}
& $1$ & $2$ \\
massive &  &  &  &  \\
& $\bar{F}^{S}$ & $\overline{%
\begin{tabular}{|c|c|}
\hline
$\ \ $&$\ \ $  \\ \hline
\end{tabular}
}$ & $1$ & $-2$ \\
&  &  &  &  \\ \cline{1-3}\cline{2-5}
&  &  &  &  \\
& $\bar{F}^{A}$ & $\overline{%
\begin{tabular}{|c|}
\hline
$\ \ $\\ \hline
$\ \ $\\ \hline
\end{tabular}
}$ & $1$ & $-2$ \\
massless &  &  &  &  \\
& $\bar{F}_{i>N}$ & $\overline{%
\begin{tabular}{|c|}
\hline
$\ \ $\\ \hline
\end{tabular}
}$ & $\overline{
\begin{tabular}{|c|}
\hline
$\ \ $\\ \hline
\end{tabular}
}$ & $-1$
\end{tabular}
\end{equation}
This breaking pattern gives $N^{2}+8N$ Goldstone bosons, $N^{2}-1$ of which
are eaten by the gauge bosons. So only $8N+1$ remain as part of the massless
spectrum along with the massless fermions. The global anomalies are again
matched by this spectrum. Those associated with the unbroken group $%
SU^{\prime }(N)\times SU(4)\times U_{1}^{\prime }(1)$ are matched by the
massless fermions, while those associated with the broken global generators
are matched by the Goldstone bosons. Since the Goldstone bosons do not
couple singly to the massless fermions (no dimension-four operators), the
effective zero-mass theory is free at low energies.

 It follows that the thermodynamic free energy may be computed at
$T
\rightarrow 0$ by counting the degrees of freedom. The result is
\begin{equation}
f_{IR}^{Higgs}(p=0) = (8N+1)+\frac{7}{4}[\frac{1}{2}N(N-1)+4N] \ ,
\end{equation}
where the superscript indicates that the gauge symmetry is
(partially) broken. Just as in the case of the symmetric phase, the
inequality Eq.~(\ref{eq:ineq}) is satisfied: $f_{IR}^{Higgs}(p=0)<
f_{UV}(p=0)$.

As an aside, we note that according to the idea of complementarity
this low energy phase may be thought of as having
arisen from confining gauge forces rather than the Higgs
mechanism~\cite{kp,ds}. Confinement then would partially break the
global symmetry to the above group forming the necessary Goldstone
bosons. It would also produce gauge singlet massless composite
fermions to replace precisely the massless elementary fermions in
the above table.

We have identified two possible phases of this theory consistent
with global anomaly matching and the inequality
Eq.~(\ref{eq:ineq}). One confines and breaks no symmetries. The
other breaks the chiral symmetry according to Eq.~(\ref{rim}). For
any finite value of $N$, $f_{IR}^{sym}(p=0)< f_{IR}^{Higgs}(p=0)$.
The symmetric phase is thus favored if the number of degrees of
freedom, or the entropy of the system near freeze-out, is
minimized. In the limit $N\rightarrow
\infty$, the Goldstone bosons do not contribute to leading order,
and $f_{IR}^{sym}(p=0)
\rightarrow f_{IR}^{Higgs}(p=0)$. We return to a discussion
of the infinite $N$ limit after describing the general ($p > 0$)
model.

What about other symmetry breaking phases of the $p=0$ theory
corresponding to bilinear condensate formation? In addition to
$S\bar{F}$ condensates, there are also $SS$ and $\bar{F}\bar{F}$
possibilities. Several of these correspond to attractive channels,
although not maximally attractive, due to gluon exchange. We have
considered all of them for the case $N=3$, and have shown that the
effective low energy theory is infrared free and that the number of
low energy degrees of freedom is larger than the symmetric phase.
Whether this is true for symmetry breaking patterns corresponding
to bilinear condensate formation for general $N$ remains to be
seen. Higher dimensional condensate formation is yet to be studied
for any of these choices.

\subsection{The General Case}

We next consider the full range of $p$ allowed by asymptotic
freedom: $0 < p< (9/2)N-3$. For $p$ near $(9/2)N - 3$, an infrared
stable fixed point exists, determined by the first two terms in the
$\beta$ function. This can be arranged by taking both $N$ and $p$
to infinity with the difference $(9/2)N - p$ fixed, or at finite
$N$ by continuing to nonintegral $p$. The infrared coupling is then
weak and the theory neither confines nor breaks symmetries. The
fixed point leads to an approximate, long-range conformal symmetry.
As $p$ is reduced, the screening of the long range force decreases,
the coupling increases, and confinement and/or symmetry breaking
set in. We consider three strong-coupling possibilities, each
consistent with global anomaly matching.

\subsubsection{Confinement with no symmetry Breaking}

It was observed by Bars and Yankielowicz~\cite{by} that
confinement without chiral symmetry breaking is consistent with
global anomaly matching provided that the spectrum of the theory
consists of massless composite fermions transforming under the
global symmetry group as follows:

\begin{equation}
\begin{tabular}{cccccc}
& $[SU(N)]$ & $SU(N+4+p)$ & $SU(p)$ & $U_{1}(1)$ & $U_{2}(1)$
\\ &  &  &  &  &  \\ $\bar{F}S\bar{F}$ & $1$ & $\overline{%
\begin{tabular}{|l|}
\hline
$\ \ $\\ \hline
$\ \ $\\ \hline
\end{tabular}
}$ & $1$ & $-N$ & $0$ \\
&  &  &  &  &  \\
$\bar{F}^{+}S^{+}F$ & $1$ & $
\begin{tabular}{|l|}
\hline
$\ \ $\\ \hline
\end{tabular}
$ & $
\begin{tabular}{|l|}
\hline
$\ \ $\\ \hline
\end{tabular}
$ & $N$ & $-N$ \\
&  &  &  &  &  \\
$F^{+}SF^{+}$ & $1$ & $1$ & $\overline{%
\begin{tabular}{|c|c|}
\hline
$\ \ $&$\ \ $  \\ \hline
\end{tabular}
}$ & $-N$ & $2N$%
\end{tabular}
\end{equation}

The effective low-energy theory is free. In Ref.~\cite{acss}, the
thermodynamic free energy for this phase was computed, giving
\begin{equation}
f_{IR}^{sym}=\frac{7}{4}[\frac{1}{2}(N+4+p)(N+3+p)+p(N+4+p)+\frac{1}{2}
p(p+1)] \ .
\end{equation}
The inequality $f_{IR}^{sym} < f_{UV}$ was then invoked to argue
that this phase is possible only if $p$ is less than a certain
value (less than the asymptotic freedom bound). For large $N$, the
condition is $p<(15/14)^{1/2}N$.

\subsubsection{Chiral symmetry breaking}

Since this theory is vector-like with respect to the $p$
$F$-$\bar{F}$ pairs, it may be anticipated that these pairs
condense according to
\begin{equation}
\overline{
\begin{tabular}{|c|}
\hline
$\ \ $\\ \hline
\end{tabular}
}\times
\begin{tabular}{|c|}
\hline
$\ \ $\\ \hline
\end{tabular}
\rightarrow 1\ ,
\end{equation}
leading to a partial breaking of the chiral symmetries. The gauge-singlet
bilinear condensate (fermion mass) is of the form
\begin{equation}
\varepsilon ^{\gamma \delta }F_{\gamma }^{a,i}\bar{F}_{a,N+4+i,\delta}\ ,
\label{chiral breaking}
\end{equation}
where $i=1,...,p$.

This leads to the symmetry breaking pattern $SU(N+4+p)\times SU(p)\times
U_{1}(1)\times U_{2}(1)\rightarrow SU(N+4)\times SU_{V}(p)\times
U_{1}^{\prime }(1)\times U_{2}^{\prime }(1)$, producing $2pN+p^{2}+8p$ gauge
singlet Goldstone bosons. The $U^{\prime }(1)^{\prime}s$ are combinations of
the $U(1)^{\prime}s$ and the broken generator of $SU(N+4+p)$
\begin{equation}
Q_{(N+4+p)}=\left(
\begin{tabular}{ccc|ccc}
$-p$ &  &  &  &  &  \\
& $\ddots $ &  &  &  &  \\
&  & $-p$ &  &  &  \\ \hline
&  &  & $N+4$ &  &  \\
&  &  &  & $\ddots $ &  \\
&  &  &  &  & $N+4$%
\end{tabular}
\right).
\end{equation}

At this stage, the remaining massless theory is the $p=0$ theory described
above, together with the $2pN+p^{2}+8p$ gauge-singlet Goldstone bosons.
Since the Goldstone bosons are associated with the broken symmetry, there
will be no dimension-four interactions between them and the $p=0$ theory.
This theory may therefore be analyzed at low energies by itself, leading to
the possible phases described above. Two possible phases of the $p=0$ theory
were discussed in detail. One corresponds to confinement and massless
composite fermion formation with no chiral symmetry breaking. For the
general theory, this corresponds to

\begin{itemize}
\item  Partial chiral symmetry breaking but no gauge symmetry breaking. The
vector-like $p$ pairs of $F$ and $\bar{F}$ condense, and others
form composite fermions.
\end{itemize}

The massless spectrum consists of the $2pN+p^{2}+8p$ Goldstone
bosons together with the $(N+4)(N+3)/2$ composite fermions of the
$p=0$ sector. All are confined. The final global symmetry is
$SU(N+4)\times SU_{V}(p)\times U_{1}^{\prime }(1)\times
U_{2}^{\prime }(1)$. Global anomalies are matched partially by the
massless composites and partially by the Goldstone bosons. Since
both theories are infrared free, the free energy may be computed in
the $T \rightarrow 0$ limit to give

\begin{equation}
f_{IR}^{brk+sym}=(2pN+p^{2}+8p)+\frac{7}{4}[\frac{1}{2}(N+4)(N+3)]
\ .
\end{equation}
The inequality Eq.~(\ref{eq:ineq}) thus allows this phase for $p$
less than a certain value below the asymptotic freedom bound but
above the value at which the symmetric phase becomes possible. For
large N, the limit is $p/N$ less than $\simeq 2.83$.

\bigskip

The other phase of the $p=0$ theory considered above, corresponds
to the MAC for symmetry breaking and the Higgsing of the gauge
group with a further breaking of the chiral symmetry. For the
general theory ($p > 0$), it leads to

\begin{itemize}
\item Further chiral symmetry breaking and gauge symmetry breaking.
\end{itemize}

The final global symmetry is $SU^{\prime}(N) \times SU(4) \times
SU_{V}(p)\times U_{1}^{\prime }(1)\times U_{2}^{\prime }(1)$. The
massless spectrum consists of the $2pN+p^{2}+8p$ Goldstone bosons
associated with the $p$ $F$-$\bar{F}$ pairs, together with the
$8N+1$ Goldstone bosons and $N(N+1)/2 + 4N$ massless elementary
fermions of the $p=0$ sector. Global anomalies are matched
partially by Goldstone bosons and partially by the remaining
massless fermions. The effective low energy theories are infrared
free, and we have

\begin{eqnarray}
f_{IR}^{brk+Higgs} &=&(2pN+p^{2}+8p)+(8N+1)  \nonumber \\
&&+\frac{7}{4}[\frac{1}{2}N(N-1)+4N] \ .
\end{eqnarray}

\begin{figure}[htb]
\center{
\epsfysize=8.0 cm
\leavevmode \epsfbox{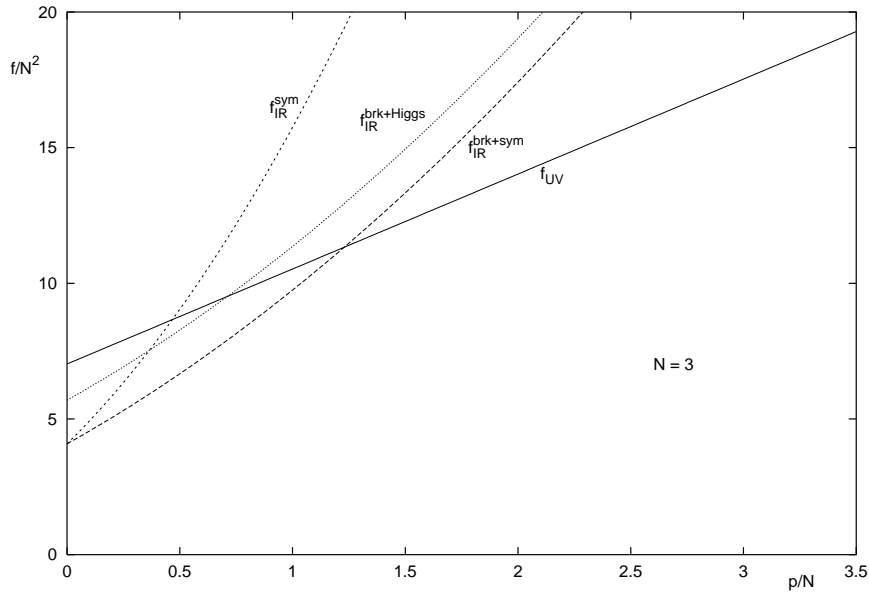}
\caption{BY
Model: Degree of freedom count $f$ (normalized to $N^2$) for
different phases as function of the number $p$ of $F-\bar{F}$ pairs
for the choice $N=3$. $f_{IR}^{sym}$ indicates confinement with
intact chiral global symmetry while $f_{IR}^{brk+sym}$ indicates
confinement with partial chiral symmetry breaking.
$f_{IR}^{brk+Higgs}$ indicates partial chiral symmetry breaking
with gauge symmetry breaking. $f_{UV}$ counts the underlying
degrees of freedom. As $N$ increases, the $f_{IR}^{brk+sym}$ and
$f_{IR}^{brk+Higgs}$ curves approach each other.}
\label{BYfp}}
\end{figure}

\bigskip

Three possible phases of the general Bars-Yankielowicz model have
now been identified. In Fig.~\ref{BYfp}, we summarize the
computation of $f_{IR}/N^{2}$ for each phase and compare with
$f_{UV}$ for the choice $N=3$. Other choices are qualitatively the
same. Each phase satisfies the inequality Eq.~(\ref{eq:ineq}) for
$p/N$ small enough. As $p$ is reduced, the first phase allowed by
the inequality corresponds to confinement with condensation of the
$p$ fermions in the real representation of the gauge group and the
breaking of the associated chiral symmetry, along with unbroken
chiral symmetry and massless composite fermion formation in the
$p=0$ sector. The degree of freedom count is denoted by
$f^{brk+sym}_{IR}$.

 The two other phases are also allowed by the inequality as $p$ is
reduced further. But for any finite value of $N$ and for any value
of $p>0$, the curve for $f_{IR}^{brk+sym}$ is the lowest of the
$f_{IR}$ curves. Thus the lowest infrared degree-of-freedom count
corresponds to a complete breaking of the chiral symmetry
associated with the $p$ $F$-$\bar{F}$ pairs (the vector- like part
of the theory with the fermions in a real representation of the
gauge group), and no breaking of the chiral symmetry associated
with the $p=0$ sector (the part of the theory with the fermions in
complex representations).

It is instructive to examine this model in the infinite $N$ limit.
If the limit is taken with $p/N$ fixed, the curves for $
f_{IR}^{brk+sym}/N^{2}$ and $f_{IR}^{brk+Higgs}/N^{2}$ become
degenerate for all values of $p/N$, and are below the curve for
$f_{IR}^{sym}$. If the limit $N \rightarrow \infty$ is taken with
$p$ fixed, all the curves become degenerate, and the phases are not
distinguished by the number of degrees of freedom. The authors of
Ref.~\cite{eppz} analyzed the model in the $N \rightarrow
\infty$ limit with confinement assumed and
 noted that the $U_{1}(1)$ symmetry cannot break
because no appropriate order parameter can form in this limit. This
is consistent with the above discussion since each of the phases
preserves the $U_{1}(1)$ for any $N$.

As with the $p=0$ theory, there are other possible symmetry
breaking phases corresponding to bilinear condensate formation.
Some of these are attractive channels, although not maximally
attractive, due to gluon exchange. We have considered several
possibilities. Each leads to an effective low energy theory that is
infrared free, and each gives a larger value of $f_{IR}$ than the
phase corresponding to the lowest curve in Fig.~\ref{BYfp}:
complete breaking of the chiral symmetry associated with $p$
additional $F$-$\bar{F}$ pairs and no breaking of the chiral
symmetry associated with the sector of the theory with the fermions
in complex representations.

Symmetry breaking by higher dimensional condensate formation is yet
to be considered, and we have nothing to say about possible
strongly coupled infrared phases such as a strongly coupled
nonabelian Coulomb phase.

\section{ The Generalized Georgi-Glashow (GGG) Model}
\label{tre}

This model is similar to the BY model just considered. It is an
$SU(N\geq 5)$ gauge theory, but with fermions in the
anti-symmetric, rather than symmetric, tensor representation. The
complete fermion content is $A=\psi _{L}^{[ab]},$ $a,b=1,\cdots
,N$; an additional $N-4+p$ fermions in the conjugate fundamental
representations: $\bar{F}_{a,i}=\psi
_{a,iL}^{c},$ $i=1,\cdots ,N-4+p$; and $p$ fermions in the fundamental
representations, $F^{a,i}=\psi _{L}^{a,i}$, $i=1,\cdots ,p$.

The global symmetry is
\begin{equation}
G_{f}=SU(N-4+p)\times SU(p)\times U_{1}(1)\times U_{2}(1) \ .
\end{equation}
where the two $U(1)$'s are anomaly free. With respect to this symmetry, the
fermion content is
\begin{equation}
\begin{tabular}{cccccc}
& $[SU(N)]$ & $SU(N-4+p)$ & $SU(p)$ & $U_{1}(1)$ & $U_{2}(1)$ \\
&  &  &  &  &  \\
$A$ &
\begin{tabular}{|c|}
\hline
$\ \ $\\ \hline
$\ \ $\\ \hline
\end{tabular}
& $1$ & $1$ & $N-4$ & $2p$ \\
&  &  &  &  &  \\
$\bar{F}$ & $\overline{%
\begin{tabular}{|c|}
\hline
$\ \ $\\ \hline
\end{tabular}
}$ & $\overline{%
\begin{tabular}{|c|}
\hline
$\ \ $\\ \hline
\end{tabular}
}$ & $1$ & $-(N-2)$ & $-p$ \\
&  &  &  &  &  \\
$F$ &
\begin{tabular}{|c|}
\hline
$\ \ $\\ \hline
\end{tabular}
& $1$ &
\begin{tabular}{|c|}
\hline
$\ \ $\\ \hline
\end{tabular}
& $N-2$ & $-(N-p)$%
\end{tabular}
\end{equation}

\bigskip

For the $\beta $-function, we have $\beta _{1}=3N+2-(2/3)p$ and $\beta
_{2}=(1/4)\{13N^{2}+30N+1+12/N-2p((13/3)N-1/N)\}$. Thus the theory is
asymptotically free if
\begin{equation}
p<(9/2)N+3 \ .
\end{equation}
We restrict $p$ so that this condition is satisfied. Because of
asymptotic freedom, the thermodynamic free-energy may be computed
in the $T\rightarrow \infty $ limit. We have
\begin{equation}
f_{UV}=2(N^{2}-1)+\frac{7}{4}{\large
[}\frac{N(N-1)}{2}+(N-4)N+2pN{\Large ]}\ .
\end{equation}
As with the BY model, we first discuss the $p=0$ theory and then
consider the general case.

\subsection{The $p=0$ Case}

The global symmetry group is $G_{f}=SU(N-4)\times U_{1}(1)$. The
theory is strongly coupled at low energies, so it is expected
either to confine or to break some of the symmetries, consistent
with global anomaly matching~\cite{thooft}.

In the case of complete confinement and unbroken symmetry, to
satisfy global anomaly matching the massless spectrum consists of
gauge singlet composite fermions $\bar{F}_{\{i}A\bar{F}_{j\}}$
transforming according to the symmetric second-rank tensor
representation of $SU(N-4)$ with charge $-N$ under the $ U_{1}(1)$
global symmetry~\cite{by}. The composites are noninteracting in the
infrared. Therefore the thermodynamic free energy may be computed
in the limit $T\rightarrow 0$. Enumerating the degrees of freedom
gives
\begin{equation}
f_{IR}^{sym}(p=0)=\frac{7}{4}\frac{(N-4)(N-3)}{2} \ .
\end{equation}
Clearly $f_{IR}^{sym}(p=0)<f_{UV}(p=0)$, satisfying the inequality
Eq.~(\ref{eq:ineq}).

We next consider symmetry breaking due to bilinear condensate
formation by first examining the maximally attractive channel
\cite{peskin}:
\begin{equation}
\begin{tabular}{|c|}
\hline
$\ \ $\\ \hline
$\ \ $\\ \hline
\end{tabular}
\times \overline{
\begin{tabular}{|c|}
\hline
$\ \ $\\ \hline
\end{tabular}
}\rightarrow
\begin{tabular}{|c|}
\hline
$\ \ $\\ \hline
\end{tabular}
,
\end{equation}
leading to the formation of the $A\bar{F}$ condensate
\begin{equation}
\varepsilon ^{\gamma \delta }A_{\gamma }^{ai}
\bar{F}_{a,i, \delta } \ ,
\label{AF}
\end{equation}
where $\gamma ,\delta =1,2$ are spin indices, $a=1,\cdots ,N$, is a
gauge index and $i=1,\cdots ,N-4$ is a flavor index. This
condensate breaks the $ U_{1}(1)$ symmetry and breaks the gauge
symmetry $SU(N)$ to $SU(4)$. The broken gauge subgroup $SU(N-4)$
combines with the flavor group, leading to a new global symmetry
$SU^{\prime }(N-4)$, while the broken gauge $SU(N)$ generator
\[
Q_{(N)}=\left(
\begin{tabular}{ccc|ccc}
$4$ &  &  &  &  &  \\
& $\ddots $ &  &  &  &  \\
&  & $4$ &  &  &  \\ \hline
&  &  & $4-N$ &  &  \\
&  &  &  & $\ddots $ &  \\
&  &  &  &  & $4-N$%
\end{tabular}
\right) \ ,
\]
combines with $U_{1}(1)$ to form a residual global symmetry
$U^{\prime }(1)$. The remaining symmetry is thus $[SU(4)]\times
SU^{\prime }(N-4)\times U_{1}^{\prime }(1)$. \ All Goldstone bosons
are eaten by gauge bosons.

We have
\begin{equation}
\bar{F}_{a,i}=\left(
\begin{tabular}{c}
$\bar{F}_{j,i}\rightarrow \bar{F}_{[j,i]}+\bar{F}_{\{j,i\}}$ \\
\hline $\bar{F}_{c,i}$
\end{tabular}
\right)
\end{equation}
and
\begin{equation}
A^{ab}=\left(
\begin{tabular}{c|c}
$A^{ij}$ & $A^{ic}$ \\ \hline & $A^{cd}$
\end{tabular}
\right) ,
\end{equation}
where $a,b=1,\cdots ,N$, \ $i,j=1,\cdots ,N-4$, and $c,d=N-3,\cdots
,N$. The $A\bar{F}$ condensate pairs $\bar{F}_{[j,i]}$ with
$A^{ij}$ and $\bar{F}
_{c,i}$ with $A^{ic}$. This leaves only $A^{cd}$,  which is neutral under $
U^{\prime }(1)$, as the fermion content of the $SU(4)$ gauge
theory.

This $SU(4)$ theory is also strongly coupled in the infrared and we
expect it to confine. The most attractive channel for condensate
formation, for example, is
\begin{equation}
\begin{tabular}{|c|}
\hline
$\ \ $\\ \hline
$\ \ $\\ \hline
\end{tabular}
\times
\begin{tabular}{|c|}
\hline
$\ \ $\\ \hline
$\ \ $\\ \hline
\end{tabular}
\rightarrow
\begin{tabular}{|c|}
\hline
$\ \ $\\ \hline
$\ \ $\\ \hline
$\ \ $\\ \hline
$\ \ $\\ \hline
\end{tabular}
=1 \ ,
\end{equation}
leading to the bilinear condensate
\begin{equation}
\varepsilon ^{\gamma \delta }A_{\gamma }^{ab}A_{\delta }^{cd}\varepsilon
_{1\cdots (N-4)abcd} \ ,  \label{AA}
\end{equation}
a singlet under the gauge group. Thus, in the infrared, the only
massless fermions are the $\bar{F}_{\{j,i\}}'s$ in the symmetric
two-index tensor representation of $SU^{\prime }(N-4)$.
Interestingly, the massless fermion content and the low energy
global symmetry are precisely the same for the symmetric and Higgs
phases. Therefore,
\begin{equation}
f_{IR}^{higgs}(p=0)= f_{IR}^{sym}(p=0)
= \frac{7}{4}[\frac{1}{2}(N-4)(N-3)]\ .
\end{equation}
The fermions are composite in the first case and elementary in the
second. This is another example of the complementarity
idea~\cite{ds}.  While the two phases are not distinguished by the
low energy considerations used here, they {\it are} different
phases. However, other ideas involving energies on the order of the
confinement and/or breaking scales will have to be employed to
distinguish them.

A general study of the phases of chiral gauge theories should
include higher dimensional as well as bilinear condensate
formation. We have done this for one case, the $p=0$ $SU(N=5)$
model, which possesses only a $U(1)$ global symmetry. Among the
various phases that may be considered is one that confines but
breaks the global $U(1)$. This corresponds to the formation of
gauge invariant higher dimensional condensates, for example of the
type $\left(\bar{F}A\bar{F}\right)^2$. There is no bilinear
condensate for this breaking pattern. Global anomaly matching is
satisfied by the appearance of a single massless Goldstone boson
and no other massless degrees of freedom. This phase clearly
minimizes the degree of freedom count (the entropy near
freeze-out), among the phases described by infrared free effective
theories. The unbroken phase, by contrast, must include a massless
composite fermion for anomaly matching, and therefore gives a
larger $f_{IR}$. This suggests that higher dimensional condensate
formation may indeed be preferred in this model. It will be
interesting to study this possibility in more detail and to see
whether higher dimensional condensate formation plays an important
role in the larger class of chiral theories considered here and in
other theories.

\subsection{The General Case}

The full range of $p$ allowed by asymptotic freedom may be
considered just as it was for the BY model. For $p$ near
$(9/2)N+3$, an infrared stable fixed point exists, determined by
the first two terms in the $\beta $ function. The infrared coupling
is then weak and the theory neither confines nor breaks symmetries.
As $p$ decreases, the coupling strengthens, and confinement and/or
symmetry breaking set in. We consider two possibilities consistent
with global anomaly matching.

\subsubsection{Confinement with no symmetry breaking}

It is known~\cite{by} that confinement without chiral symmetry
breaking is consistent with global anomaly matching provided that
the spectrum of the theory consists of gauge singlet massless
composite fermions transforming under the global symmetry group as
follows:
\begin{equation}
\begin{tabular}{cccccc}
& $[SU(N)]$ & $SU(N-4+p)$ & $SU(p)$ & $U_{1}(1)$ & $U_{2}(1)$ \\ &
&  &  &  &  \\ $\bar{F}A\bar{F}$ & $1$ & $\overline{
\begin{tabular}{|c|c|}
\hline
$\ \ $&$\ \ $  \\ \hline
\end{tabular}
}$ & $1$ & $-N$ & $0$ \\
&  &  &  &  &  \\
$\bar{F}^{+}A^{+}F$ & $1$ & $
\begin{tabular}{|l|}
\hline
$\ \ $\\ \hline
\end{tabular}
$ & $
\begin{tabular}{|l|}
\hline
$\ \ $\\ \hline
\end{tabular}
$ & $N$ & $-N$ \\
&  &  &  &  &  \\
$F^{+}AF^{+}$ & $1$ & $1$ & $\overline{%
\begin{tabular}{|l|}
\hline
$\ \ $\\ \hline
$\ \ $\\ \hline
\end{tabular}
}$ & $-N$ & $2N$%
\end{tabular}
\end{equation}
The effective low energy is free. Thus the thermodynamic free
energy may be computed in the limit $T \rightarrow 0$ to give
\begin{equation}
f_{IR}^{sym}=\frac{7}{4}[\frac{1}{2}(N-4+p)(N-3+p)+p(N-4+p)+\frac{1}{2}
p(p-1)] \ .
\end{equation}
The inequality Eq.~(\ref{eq:ineq}) allows this phase when $p/N$ is
less than $\simeq 2.83$, for large $N$.

\subsubsection{Chiral symmetry breaking}

As in the BY model it may be expected that the fermions in a real
representation of the gauge group (the $p$ $F$-$\bar{F}$ pairs)
will condense in the pattern
\begin{equation}
\overline{
\begin{tabular}{|c|}
\hline
$\ \ $\\ \hline
\end{tabular}
}\times
\begin{tabular}{|c|}
\hline
$\ \ $\\ \hline
\end{tabular}
\rightarrow 1.
\end{equation}
The gauge-singlet bilinear condensate (fermion mass) is of the form
\begin{equation}
\varepsilon ^{\gamma \delta }F_{\gamma }^{a,i}\bar{F}_{a,N-4+i,\delta} \ ,
\end{equation}
where $i=1,...,p$, leading to the symmetry breaking pattern
\begin{equation}
\begin{tabular}{c}
$SU(N-4+p)\times SU(p)\times U_{1}(1)\times U_{2}(1)$ \\
$\rightarrow SU(N-4)\times SU_{V}(p)\times U_{1}^{\prime }(1)\times
U_{2}^{\prime }(1),$
\end{tabular}
\end{equation}
and producing $2pN+p^{2}-8p$ gauge singlet Goldstone bosons.

The $U^{\prime }(1)^{\prime }s$ are combinations of the
$U(1)^{\prime }s$ and the broken generator of $SU(N-4+p)$
\begin{equation}
Q_{(N-4+p)}=\left(
\begin{tabular}{ccc|ccc}
$-p$ &  &  &  &  &  \\
& $\ddots $ &  &  &  &  \\
&  & $-p$ &  &  &  \\ \hline
&  &  & $N-4$ &  &  \\
&  &  &  & $\ddots $ &  \\
&  &  &  &  & $N-4$%
\end{tabular}
\right)\ .
\end{equation}

The remaining massless theory is the $p=0$ theory described above,
together with the $2pN+p^{2}-8p$ gauge-singlet Goldstone bosons.
Since the Goldstone bosons are associated with the broken symmetry,
there will be no dimension-four (Yukawa) interactions between them
and the $p=0$ fields. The $p=0$ theory may therefore be analyzed by
itself, leading to the possible phases described above. Two phases
were considered, one symmetric and the other broken by the
maximally attractive bilinear condensate, and they were seen to
lead to identical low energy theories.

Thus, in either case, the degree-of-freedom count for the general
theory, corresponding to the breaking of the chiral symmetry
associated with the $p$ $F$-$\bar{F}$ pairs, gives
\begin{equation}
f_{IR}^{brk}=(2pN+p^{2}-8p)+\frac{7}{4}[\frac{1}{2}(N-4)(N-3)] \ .
\end{equation}

\begin{figure}[htb]
\center{
\epsfysize=8.0 cm
\leavevmode \epsfbox{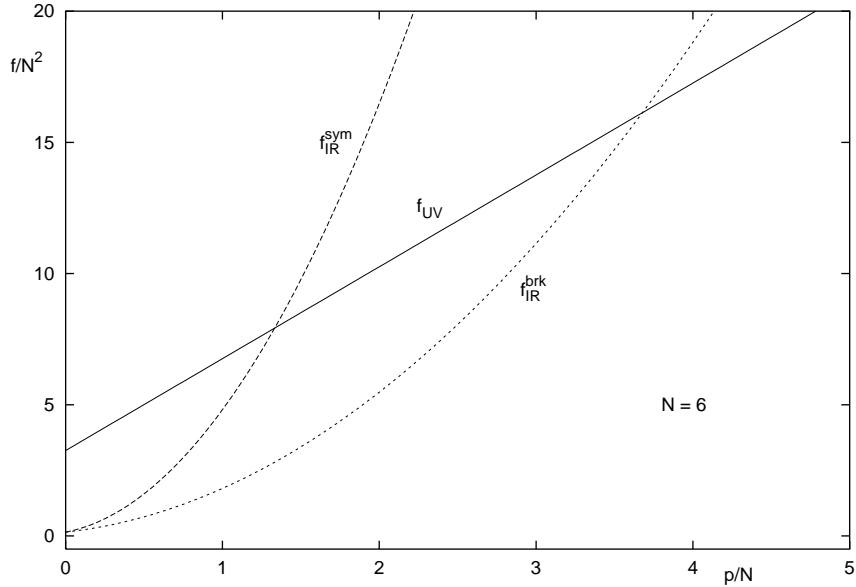}
\caption{ GGG Model:
Degree of freedom count $f$ (normalized to $N^2$) for different
phases as function of the number $p$ of $F-\bar{F}$ pairs for the
choice $N=6$. Other choices are qualitatively the same.
$f_{IR}^{sym}$ indicates confinement with intact chiral global
group while $f_{IR}^{brk}$ indicates either confinement or gauge
symmetry breaking, with partial chiral symmetry breaking. $f_{UV}$
counts the underlying degrees of freedom.}
\label{GGGfp}}
\end{figure}

\bigskip

  To summarize, two possible phases of the general GGG model have been
considered. In Fig.~\ref{GGGfp}, we plot the two computations of
$f_{IR}/N^{2}$ for the choice $N=6$ and compare them with $f_{UV}$.
Each phase satisfies the inequality Eq.~(\ref{eq:ineq}) for $p$
below some value. As $p$ is reduced, the first phase allowed by the
inequality corresponds to partial chiral symmetry breaking. For any
$p$, $f_{IR}^{brk}$ is the lower of the $f_{IR}$ curves. Thus the
lower infrared degree-of-freedom count corresponds to a complete
breaking of the chiral symmetry associated with $p$ additional
$F$-$\bar{F}$ pairs (the vector like part of the theory) and no
breaking of the chiral symmetry associated with the fermions in a
complex representation of the gauge group. Whether the latter
behavior is due to confinement or the Higgsing of the gauge group
has not been determined. These conclusions remain valid in the
infinite $N$ limit with $p/N$ fixed. If the limit $N
\rightarrow \infty$, is taken with $p$ fixed, the two curves become degenerate.

As we have already noted, for the $p=0$ $SU(5)$ theory
there is a still lower degree of freedom count  when higher
dimensional condensates are considered. This will therefore also be
true of the general-$p$ case for $SU(5)$. This even lower count
corresponds to complete confinement along with breaking of the
chiral symmetry associated with the $p$ $F$-$\bar{F}$ pairs {\it
and} breaking of the remaining global $U(1)$ symmetry. It will be interesting
to see whether a preference for this phase can be confirmed by
a dynamical study of this model and whether
similar higher dimensional condensate formation is favored in a
more general class of models.

\bigskip

\section{Two Chiral SUSY Models}
\label{quattro}

 Although this paper is devoted principally to non-SUSY chiral models,
 we briefly describe two chiral SUSY models: the supersymmetric
generalization of the one generation $SU(5)$ Georgi-Glashow model
\cite{ADSSB} and the related ($3-2$) model (see \cite{pt} for a
review of this model and relevant references).

The $SU(5)$ model contains a single antisymmetric tensor chiral
superfield $A$ and an antifundamental chiral superfield $\bar{F}$.
The vector superfield $W_{\alpha}$ includes the standard vector
boson and the associated gluino in the adjoint representation of
$SU(5)$. The global symmetry is the anomaly-free $U_R(1)\times
U_A(1)$, and the charge assignments are:

\begin{equation}
\begin{tabular}{cccccc}
& $[SU(5)]$ & $U_{R}(1)$ & $U_{A}(1)$ \\ & & & & &
\\ $A$ &
\begin{tabular}{|c|} \hline $\ \ $\\ \hline$\ \ $ \\ \hline \end{tabular}
& $-1$ & $-1$ \\ &  &  &  &  &  \\ $\bar{F}$ & $\overline{%
\begin{tabular}{|c|}
\hline
$\ \ $\\ \hline
\end{tabular}
}$ &
 $+ 9$ & $+3$ \\ &  &  &  &  &  \\ $W_{\alpha}$ &
$Adj$ & $-1$ & $ 0$%
\end{tabular}
\end{equation}

\bigskip
A special feature of this model is that the classical vacuum is
unique. The absence of flat directions is due to the fact that
there exists no holomorphic gauge invariant polynomial constructed
out of the supersymmetric fields. This feature guarantees that when
comparing phases through their degree-of-freedom count, we know
that we are considering a single underlying theory. By contrast, in
SUSY gauge theories with flat directions, non-zero condensates
associated with the breaking of global symmetries correspond to
different points in moduli space and therefore to different
theories.

This model was studied long ago \cite{ADSSB} and various possible phases
were seen to be consistent with global anomaly matching. One
preserves supersymmetry along with the global symmetries. This
requires composite massless fermions to saturate the global
anomalies. It was shown that there are several, rather complicated,
solutions, with at least five Weyl fermions (which for
supersymmetry to hold must be cast in five chiral superfields). The
charge assignments for one of them is \cite{ADSSB}:
\begin{equation}
(-5,-26),~(5,20),~(5,24),~(0,-1),~(0,9) \ ,
\end{equation}
where the first entry is the $U_A(1)$ charge and the second is the
$U_R(1)$ charge of each chiral superfield.

Other possibilities are that SUSY breaks with the global symmetries
unbroken or that one or both of the global symmetries together with
supersymmetry break spontaneously. It is expected \cite{ADSSB} that
in a supersymmetric theory without classical flat directions, the
spontaneous breaking of global symmetries also signals spontaneous
supersymmetry breaking. In these cases, the only massless fields
will be the Goldstone boson(s) associated with the broken global
symmetries and/or some massless fermions transforming under the
unbroken chiral symmetries, together with the Goldstone Weyl
fermion associated with the spontaneous supersymmetry breaking.

In Reference~\cite{ADSSB} it was suggested on esthetic grounds that
the supersymmetric solution seems less plausible. Additional
arguments that supersymmetry is broken are based on investigating
correlators in an instanton background \cite{MV}. However a firm
solution to this question is not yet available.

Since all the above phases are non interacting in the infrared we
may reliably compute $f_{IR}$ and note that the phase that
minimizes the degree-of-freedom count is the one that breaks
supersymmetry and both of the global symmetries. This phase
consists of two $U(1)$ Goldstone bosons and a single Weyl Goldstino
associated with the breaking of SUSY. Thus $f_{IR} = 15/4$. SUSY
preserving phases and those that leave one or both of the $U(1)'s$
unbroken lead to more degrees of freedom.  It will be interesting to
see whether further dynamical studies confirm that the
maximally broken phase is indeed preferred

This phase is similar to the minimal-$f_{IR}$ phase in the
nonsupersymmetric $SU(5)$ model in that both correspond to
higher dimensional condensate formation. In the SUSY case, one can
construct two independent order parameters. The one for $U_R(1)$ is
the gluino condensate (scalar component of the chiral superfield
$W^{\alpha} W_{\alpha}$) while the one for $U_A(1)$ can be taken to
be the scalar component of the chiral superfield
$\bar{F}_a\bar{F}_b A^{ac} (W^{\alpha} W_{\alpha})^b_c$.

Finally we comment on a well known and related chiral model for
dynamical supersymmetry breaking: the ($3-2$) model. Unlike the
models considered so far, this model involves multiple couplings,
i.e. two gauge couplings and a Yukawa one. Without the Yukawa
interaction the theory posses a run-away vacuum. The model has an
$SU(3)\times SU(2)$ gauge symmetry and a $U_{Y}(1)\times U_{R}(1)$
anomaly free global symmetry. As above, the low energy phase that
minimizes the number of degrees of freedom is the one that breaks
supersymmetry along with both of the global symmetries. The
massless spectrum is the same as in the parent chiral $SU(5)$ case.
In the ($3-2$) model, however, the low energy spectrum has been
computed \cite{BPR} in a self-consistent weak-Yukawa coupling
approximation, where it was noted that the $U_{R}(1)$ breaks along
with supersymmetry, leaving intact the $U_{Y}(1)$. The spectrum
consists of two massless fermions (a Goldstino and the fermion
associated with the unbroken $U_{Y}(1)$) and the $U_{R}(1)$
Goldstone boson. If this is indeed the ground state, then the
number of infrared degrees of freedom is not minimized in this weak
coupling case.

\section{Conclusions}
\label{cinque}

 We have considered the low energy structure of two chiral gauge
 theories, the Bars-Yankielowicz (BY) model and the generalized
 Georgi-Glashow (GGG) model. Each contains a core of fermions
 in complex representation of the gauge group, along with a set
 of $p$ additional fundamental-anti-fundamental pairs. In each case,
 for $p$ near but not above the value for which asymptotic freedom is lost,
the model will have a weak infrared fixed point and exist in the
non-abelian Coulomb phase.

  As $p$ drops, the infrared coupling strengthens and one or more phase
transitions to strongly coupled phases are expected. Several
possible phases have been identified that are consistent with
global anomaly matching, and that satisfy the inequality
Eq.~(\ref{eq:ineq}) for low enough $p$. One is confinement with the
gauge symmetry and additional global symmetries unbroken. Another
is confinement with the global symmetry broken to that of the $p=0$
theory. Still another is a Higgs phase, with both gauge and chiral
symmetries broken. Both symmetry breaking phases correspond to
bilinear condensate formation. The infrared degree of freedom count
$f_{IR}$ for each of these phases is shown in Figs.~\ref{BYfp} and \ref{GGGfp},
along with the corresponding ultraviolet count $f_{UV}$.

We have suggested that at each value of $p$, these theories will
choose the phase that minimizes the degree of
freedom count as defined by $f_{IR}$, or equivalently
the phase that minimizes the entropy near freeze-out ($S(T)\approx
(2\pi^{2}/45) T^{3} f_{IR}$). As may be seen from
 Figs.~\ref{BYfp} and \ref{GGGfp}, this idea leads to the following picture. As $p$
drops below some critical value, the $p$
fundamental-anti-fundamental pairs condense at some scale
$\Lambda$, breaking the full global symmetry to the symmetry of the
$p=0$ theory and producing the associated Goldstone bosons. For the
remaining theory with fermions in only complex representations, the
phase with the global symmetry unbroken and the global anomalies
matched by massless fermions is preferred to phases with further
global symmetry breaking via bilinear condensates. We have not yet
shown that this is true relative to all bilinear condensate
formation. Also, this does not exclude the possibility that some
strongly coupled infrared phase (such as a strong non abelian
coulomb phase) leads to the smallest value for $f_{IR}$ and is
still consistent with global anomaly matching.

We extended our discussion to include general condensate formation
for one simple example, the $SU(5)$ Georgi-Glashow model with
fermions in only complex representations and a single $U(1)$ global
symmetry. This symmetry can be broken via only a higher dimensional
condensate. For this model, interestingly, we noted that the
breaking of the $U(1)$ with confinement and unbroken gauge symmetry
leads to the minimum value of $f_{IR}$ among phases that are
infrared free. This highlights the important question of the
pattern of symmetry breaking in general chiral theories (or any
theories for that matter) when arbitrary condensate formation is
considered. Higher dimensional condensates could play an important
role in the dynamical  breaking of symmetries in extensions of the
standard model \cite{HDC}. The enumeration of degrees of freedom in
the effective infrared theory is a potentially useful guide to
discriminate among the possibilities.

Finally, we commented on two supersymmetric chiral models: the
supersymmetric version of the $SU(5)$ Georgi-Glashow model and the
closely related ($3-2$) model. Both have a $U_{R}(1) \times
U_{Y}(1)$ global symmetry. In each case, the phase that minimizes
the number of massless degrees of freedom corresponds to the
breaking of SUSY and both of its global symmetries. In the case of
the ($3-2$) model, however, an analysis in the case of a weak
Yukawa coupling (see \cite{pt} for a discussion and relevant
references) leads to the conclusion that the $U_{Y}(1)$ is not
broken. If this truly represents the ground state in the case of
weak coupling, then the degree of freedom count is not the minimum
among possible phases that respect global anomaly matching.

To summarize, for the nonsupersymmetric chiral gauge theories
discussed here, we have identified a variety of possible
zero-temperature phases and conjectured that the theories will
choose from among them the one that minimizes the infrared degree
of freedom count. Whether this can be proven and whether the idea
plays a role in a wider class of theories remains to be seen.

\vskip2cm
\centerline{\bf Acknowledgments}

We thank Andrew Cohen, Erich Poppitz, Nicholas Read and Robert
Shrock for helpful discussions. The work of T.A., Z.D. and F.S. has
been partially supported by the US DOE under contract
DE-FG-02-92ER-40704.

\end{document}